\documentclass[10pt,twocolumn, conference]{IEEEtran}

\usepackage{amsmath}
\usepackage{amssymb}
\usepackage[dvips]{graphicx}
\usepackage{setspace}
\usepackage{epsfig}
\usepackage{epsfig}
\usepackage{srcltx}
\usepackage{srctex}

\newcommand{\tSNR}{\text{SNR}}
\newcommand{\E}{\mathbb{E}}
\newcommand{\figsize}{0.49}

\newtheorem{theo}{Theorem}

\begin{document}
\title{QoS Analysis of Cognitive Radio
Channels with Perfect CSI at both Receiver and Transmitter}
\author{\authorblockN{Sami Akin and
Mustafa Cenk Gursoy}
\authorblockA{Department of Electrical Engineering\\
University of Nebraska-Lincoln\\ Lincoln, NE 68588\\ Email:
samiakin@huskers.unl.edu, gursoy@engr.unl.edu}}
\date{}

\maketitle

\begin{abstract}\footnote{This work was supported by the National Science Foundation under
Grants CNS -- 0834753 and CCF -- 0917265.}
In this paper, cognitive transmission under quality of
service (QoS) constraints is studied. In the cognitive radio channel model,
it is assumed that both the secondary receiver and the secondary transmitter
know the channel fading coefficients perfectly and optimize the power adaptation policy under given constraints, depending on the channel activity of the
primary users, which is determined by channel sensing performed by the
secondary users. The transmission rates are equal to the instantaneous channel
capacity values. A state transition model with four states is constructed to
model this cognitive transmission channel. Statistical limitations on
the buffer lengths are imposed to take into account the QoS constraints.
The maximum throughput under these statistical QoS constraints
is identified by finding the effective capacity of the cognitive
radio channel. The impact upon the effective capacity of several
system parameters, including the channel sensing duration,
detection threshold, detection and false alarm probabilities, and QoS
parameters, is investigated.
\end{abstract}

\section{Introduction}
With growing demand for spectrum use in wireless
communications, efficient use of the spectrum is becoming necessary.
On the other hand, recent studies show that licensed users
are not utilizing the spectrum efficiently. This has led to much
interest in cognitive radio systems.
Since the basic principle in cognitive transmission is not to disturb the primary users of the
spectrum, it is very important to detect the activities of the
primary users and not cause any harmful interference to their
communications. When the primary users are active, the secondary
user should either avoid using the channel or spend low power not to
exceed the noise power threshold of the primary users, whereas when
the channel is free of the primary users, secondary users can
use the channel without any constraints.

Considering a model which describes the busy and idle periods of a
wireless LAN (WLAN), Geirhofer
\textit{et al.} in their studies \cite{Geirhofe} focused on dynamically sharing the spectrum in the
time-domain by exploiting the whitespace between the bursty
transmissions of a set of users, represented by an 802.11b based
WLAN. They found that Pareto distribution provides an adequate
fit for varying packet rates and they proposed a continuous-time
semi-Markov model that captures the idle periods remaining between
the bursty transmission of WLAN. The problem of maximally utilizing
the spectrum opportunities in cognitive radio networks with multiple
potential channels and the problem of optimal sensing order in
multi-channel cognitive medium access control with opportunistic
transmission are investigated in \cite{Lai}. The authors in
\cite{Srinivasa} studied the opportunistic secondary communication
over a spectral pool of two independent channels and showed that the
benefits of spectral pooling are lost in dynamic spectral
environments.

In communication systems, it is very important to satisfy certain quality of service (QoS)
constraints in order to guarantee an expected level of performance to the users. However, this is a very challenging
task in wireless systems because of the random variations in channel conditions and the resulting
random fluctuations in received power levels. Consequently, only statistical, rather than deterministic, service guarantees can most of the time be provided in wireless communications.  Note that the situation is further exasperated in cognitive wireless systems due to additional random fluctuations in the transmitted power levels, depending on the presence or absence of primary users. For instance, when there are active primary users present, secondary users either cease the transmission or transmit only at lower power levels, while they can transmit at higher power levels in the absence of active primary users.
Furthermore, cognitive radio can suffer from errors in channel
sensing in the form of false alarms. Hence, it is of interest to
analyze the performance of cognitive radio systems in the presence
of QoS constraints.

The notion of effective capacity is a useful tool to
obtain the maximum throughput levels in wireless systems under
statistical QoS constraints. The application and analysis of
effective capacity in various settings has attracted much interest.
The effective capacity is defined by Wu and Negi in \cite{Wu} as the maximum constant
arrival rate that a given time-varying service process can support
while meeting QoS requirements. The
authors in \cite{Gursoy}, employing the normalized the effective capacity as the performance metric, explored the energy efficiency in the
low-power and wideband regimes under
QoS constraints. They considered variable-rate/variable-power and
variable-rate/fixed-power transmission schemes assuming the
availability of channel side information at both transmitter and
receiver or only at the receiver. In their study \cite{Liu2}, Liu \emph{et al.} focused on the effective capacity and analyzed the
resource requirements for Markov wireless channel models while
considering fixed-rate transmission schemes and the continuous-time
Gilbert-Elliot channel with ON and OFF states. Tang and Zhang in \cite{Tang} derived the optimal power and
rate adaptation techniques that maximize the system throughput under
QoS constraints, assuming that the instantaneous channel gains are known by both the
transmitter and the receiver. In \cite{Sami}, we studied the effective capacity of
cognitive radio channels in which the cognitive radio detects the
activity of primary users and then performs the data transmission. We assumed that only the receiver has channel side
information (CSI), and the transmitter, having no information about
the fading coefficients, sends the data
at two different fixed rates with two different power levels
depending on the results of channel sensing.

In this study, we again focus on the effective capacity of cognitive
radio channels but we now study the scenario in which both
the receiver and the transmitter have perfect CSI and hence know the instantaneous values of the fading
coefficients. We investigate the performance when power and rate adaptation is employed in the system.

The
organization of the rest of the paper is as follows. In Section
\ref{sec:system}, we describe the system and the cognitive channel
model. In Section \ref{sec:sensing}, we discuss channel sensing and
provide expressions for the detection and false alarm probabilities.
In Section \ref{sec:transition}, we describe the state transition
model for the cognitive radio channel and obtain the effective
capacity expression. In Section \ref{sec:numericalresults}, we
identify the impact of channel sensing parameters, detection and
false alarm probabilities and QoS constraints on the effective
capacity through numerical analysis. Finally, in Section
\ref{sec:conclusion}, we provide conclusions.

\section{System and Cognitive Channel Model} \label{sec:system}

We consider a cognitive radio channel model in which a secondary
transmitter aims to send information to a secondary receiver with
the possibility of collision with the primary users. Initially,
secondary users perform channel sensing, and then depending on the
primary users' activity, the transmitter selects its transmission
power and rate, i.e., when the channel is busy, the symbol power is
$P_{1}(i)$ and the rate is $r_{1}(i)$, and when the channel is idle,
the symbol power is $P_{2}(i)$ and the rate is $r_{2}(i)$, where $i$ denotes the time index. We assume
that the data generated by the source is initially stored in the
data buffer before being transmitted in frames of duration $T$
seconds over the cognitive wireless channel. The discrete-time
channel input-output relation in the $i^{\text{th}}$ symbol duration
is given by
\begin{align}\label{input-out1}
&y(i)=h(i)x(i)+n(i) \quad i=1,2,\dots
\end{align}
if the primary users are absent. On the other hand, if primary users
are present in the channel, we have
\begin{align}\label{input-out2}
&y(i)=h(i)x(i)+s_{p}(i)+n(i) \quad i=1,2,\dots
\end{align}
Above, $x(i)$ and $y(i)$ denote the complex-valued channel input and
output, respectively. We assume that the bandwidth available in the
system is \textit{B} and the channel input is subject to the
following average energy constraints:
\begin{align}
&\mathbb{E}\{|x(i)|^{2}\}=\mathbb{E}\{P_{1}(i)\}/B\leq \overline{P}_{1}/B\nonumber\\
&\mathbb{E}\{|x(i)|^{2}\}=\mathbb{E}\{P_{2}(i)\}/B\leq \overline{P}_{2}/B\nonumber
\end{align}
for all $\textit{i}$, when the channel is busy and idle,
respectively. Since the bandwidth is \textit{B}, symbol rate is
assumed to be \textit{B} complex symbols per second, indicating that
the average power of the system is constrained by $\overline{P}_{1}$
or $\overline{P}_{2}$. In (\ref{input-out1}) and (\ref{input-out2}),
$h(i)$ is the independent, zero mean, circular, complex Gaussian
channel fading coefficient between the cognitive transmitter and the
receiver, and has a variance of $\sigma_{h}^{2}$. We denote the
magnitude of the fading coefficients by $z(i)=|h(i)|^{2}$. We
consider a block-fading channel model and assume that the fading
coefficients stay constant for a block of duration $T$ seconds and
change independently from one block to another independently.

In (\ref{input-out2}), $s_{p}(i)$ represents the sum of the active
primary users' faded signals arriving at the secondary receiver. In
the input-output relations (\ref{input-out1}) and
(\ref{input-out2}), $n(i)$ models the additive thermal noise at the
receiver, and is a zero-mean, circularly symmetric, complex Gaussian
random variable with variance
$\mathbb{E}\{|n(i)|^{2}\}=\sigma_{n}^{2}$ for all $i$.

\section{Channel Sensing} \label{sec:sensing}

If the transmission strategies of the primary users are not known,
energy-based detection methods are well-suited for the detection of
the activities of primary users. We assume that the first $N$
seconds of the frame duration $T$ is allocated to sense the channel.
The channel sensing can be formulated as a hypothesis testing
problem between the noise $n(i)$ and the signal $s_p(i)$ in noise.
Noting that there are $NB$ complex symbols in a duration of $N$
seconds, this can mathematically be expressed as follows:
\begin{align}\label{hypothesis}
&\mathcal{H}_{0}\quad : \quad y(i)=n(i), \quad i=1,\dots,NB\\ \nonumber
&\mathcal{H}_{1}\quad : \quad y(i)=s_p(i)+n(i), \quad i=1,\dots,NB.
\end{align}
Considering the above detection problem, the optimal Neyman-Pearson
detector is given by \cite{Poor-book} as follows:
\begin{equation}\label{Neyman-Pearson}
Y=\frac{1}{NB}\sum_{i=1}^{NB}|y(i)|^{2}\gtrless^{\mathcal{H}_{1}}_{\mathcal{H}_{0}}\gamma.
\end{equation}
where $\lambda$ is the detection threshold. We assume that $s_p(i)$
has a circularly symmetric complex Gaussian distribution with
zero-mean and variance $\sigma_{s_{p}}^{2}$. Note that this is an
accurate assumption if the signals are being received in a rich
multipath environment or the number of active primary users is
large. Moreover, if, for instance the primary users are employing
phase or frequency modulation, $s_p(i)$ in the presence of even a
single primary user in flat Rayleigh fading will be Gaussian
distributed\footnote{ Note that zero-mean, circular, complex
Gaussian distributions are invariant under rotation. For instance,
if the fading coefficient $h$ is zero-mean, circularly symmetric,
complex Gaussian distributed, then so is $he^{j\phi}$ for any random
$\phi$.}. Assuming further that $\{s_p(i)\}$ are i.i.d., we can
immediately conclude that the test statistic $Y$ is chi-square
distributed with $2NB$ degrees of freedom. In this case, the
probabilities of false alarm and detection can be established as
follows:
\begin{align}\label{false alarm}
&P_{f}=Pr(Y>\gamma|\mathcal{H}_{0})=1-P\left(\frac{NB\gamma}{\sigma_{n}^{2}},NB\right)\\
&P_{d}=Pr(Y>\gamma|\mathcal{H}_{1})=1-P\left(\frac{NB\gamma}{\sigma_{n}^{2}+\sigma_{s_{p}}^{2}},NB\right) \label{eq:probdetect}
\end{align}
where $P(x,a)$ denotes the regularized lower gamma function and is
defined as $P(x,a) = \frac{\gamma(x,a)}{\Gamma(a)}$ where
$\gamma(x,a)$ is the lower incomplete gamma function and $\Gamma(a)$
is the Gamma function.

In the above hypothesis testing problem, another approach is to
consider $Y$ as Gaussian distributed, which is accurate if $NB$ is
large \cite{liang}. In this case, the detection and false alarm
probabilities can be expressed in terms of Gaussian $Q$-functions.

\section{State Transition Model and Effective Capacity} \label{sec:transition}

\subsection{State Transition Model} \label{subsec:state}

In this paper, we assume that both the receiver and the transmitter
have perfect channel side information and hence perfectly know the
instantaneous values of $\{h[i]\}$. In Figure 1, we depict the state
transition model for cognitive radio transmission.

First, note that we have the following four possible scenarios:
\begin{enumerate}
  \item Channel is busy, detected as busy
  \item Channel is busy, detected as idle
  \item Channel is idle, detected as busy
  \item Channel is idle, detected as idle.
\end{enumerate}
In each scenario, we have one state either ON or OFF, depending on
whether or not the determined transmission rate exceeds the
instantaneous channel capacity. Note that if the channel is detected
as busy, the secondary transmitter sends the information with the
power policy, $P_1(i)$ (If $P_1(i) = 0$, the secondary transmitter
stops the transmission). Otherwise, it transmits with a different
power policy, $P_2(i)$.

\begin{figure}
\begin{center}
\includegraphics[width = \figsize\textwidth]{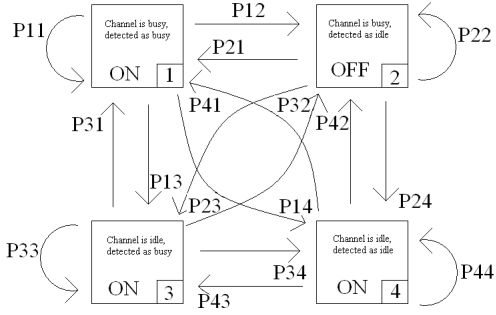}
\caption{State transition model for the cognitive radio channel. The numbered label for each state is given on the lower-right corner of the box representing the state. } \label{fig:fig1}
\end{center}
\end{figure}

Since both the transmitter and the receiver are aware of the channel
conditions, information is transmitted at rates $r_{1}(i)$ and
$r_{2}(i)$ which are equal to the estimated optimal channel
capacities, depending on the channel being detected as busy or idle,
respectively. Note that since the channel can not be sensed
correctly all the time, the transmission rates may be below or over
the actual channel capacities. Here, we can define the instantaneous
transmission powers as
$P_{1}(i)=\mu_{1}(\theta,z(i))\overline{P}_{1}$ and
$P_{2}(i)=\mu_{2}(\theta,z(i))\overline{P}_{2}$, where
$\mu_{1}(\theta,z(i))$ and $\mu_{2}(\theta,z(i))$ are the optimal
normalized power-adaptation policies for the cases when the channel is busy and
idle, respectively, and $\theta$ is $\textit{QoS}$ exponent which is
explained in the following section.

Considering the interference $s_p$ caused by the primary users as
additional Gaussian noise, we can express the instantaneous channel
capacities in the above four scenarios as follows:
\begin{align}\label{channel capacity}
&C_{1}(i)=B\log_{2}(1+\tSNR_{1}(i)z(i)) \text{ (channel busy, detected busy)}\nonumber\\
&C_{2}(i)=B\log_{2}(1+\tSNR_{2}(i)z(i)) \text{ (channel busy, detected idle)}\nonumber\\
&C_{3}(i)=B\log_{2}(1+\tSNR_{3}(i)z(i)) \text{ (channel idle, detected busy)}\nonumber\\
&C_{4}(i)=B\log_{2}(1+\tSNR_{4}(i)z(i)) \text{ (channel idle, detected idle)}.
\end{align}
where
\begin{align}
&\tSNR_{1}(i)=\frac{P_{1}(i)}{B\left(\sigma_{n}^{2}+\sigma_{s_{p}}^{2}\right)}\quad and \quad
\tSNR_{2}(i)=\frac{P_{2}(i)}{B\left(\sigma_{n}^{2}+\sigma_{s_{p}}^{2}\right)}\nonumber \\
&\tSNR_{3}(i)=\frac{P_{1}(i)}{B\sigma_{n}^{2}}\quad and \quad \tSNR_{4}(i)=\frac{P_{2}(i)}{B\sigma_{n}^{2}}.\nonumber
\end{align}
The optimal transmission rates can be expressed as follows:
\begin{align}\label{data rates}
&r_{1}(i)=B\log_{2}(1+\mu_{1}(\theta,z(i))z(i)\tSNR_{1})\nonumber\\
&r_{2}(i)=B\log_{2}(1+\mu_{2}(\theta,z(i))z(i)\tSNR_{4}).
\end{align}

Using (\ref{channel capacity}) and (\ref{data rates}), we observe that we have the ON
states when $r_{1}(i)=C_{1}(i)$, $r_{1}(i)<C_{3}(i)$ and $r_{2}(i)=C_{4}(i)$, and the OFF state
when $r_{2}(i)>C_{2}(i)$. In the OFF state, reliable communication is not achieved,
and hence, the information has to be resent. It is assumed that a simple automatic repeat
request (ARQ) mechanism is incorporated in the communication protocol to acknowledge the reception
of data and to ensure that erroneous data is retransmitted. As depicted in Fig. \ref{fig:fig1}, there
are three ON states and one OFF state.

Due to the block fading assumption, state transitions occur every $T$ seconds. When the channel is
busy and detected as busy, the probability of staying in the ON state, which is state 1 in
Fig. \ref{fig:fig1}, is expressed as $p_{11}=\rho P_{d}$, where $\rho$ is the prior probability of
channel being busy, and $P_{d}$ is the probability of detection as defined in (\ref{eq:probdetect}).
Since the state transition probabilities are independent of the original states, the other
transition probabilities become
\begin{align}\label{state transtition}
&p_{11}=p_{21}=p_{31}=p_{41}=\rho P_{d}\nonumber\\
&p_{12}=p_{22}=p_{32}=p_{42}=\rho (1-P_{d})\nonumber\\
&p_{13}=p_{23}=p_{33}=p_{43}=(1-\rho)P_{f}\nonumber\\
&p_{14}=p_{24}=p_{34}=p_{44}=(1-\rho)(1-P_{f}).
\end{align}
Then, we can easily see again due to the block-fading assumption that the $4\times4$ state
transition probability matrix can be expressed as
\begin{center}
$R = \left[
    \begin{array}{cccc}
        p_{1,1} & . & . & p_{4,1}\\
        . &  &  & .\\
        . &  &  & .\\
        p_{1,4} & . & . & p_{4,4}\\
    \end{array}
    \right]
     =\left[
    \begin{array}{cccc}
        p_{1} & . & . & p_{1}\\
        . &  &  & .\\
        . &  &  & .\\
        p_{4} & . & . & p_{4}\\
    \end{array}
\right]$.
\end{center}
Note that $R$ has a rank of 1. Note also that $r_{1}(i)(T-N)$ and $r_{2}(i)(T-N)$ bits are
transmitted and received in the states 1 and 3, and 4, respectively, while the transmitted
number of bits is assumed to be zero in state 2.

Moreover, recall that the mean transmit power is upper-bounded by $\overline{P}_{1}$
and $\overline{P}_{2}$ when the channel is busy and idle, respectively. Therefore, the optimal
power policy needs to satisfy the average power constraints:
\begin{equation}\label{mu}
\int_{0}^{\infty}\mu_{1}(\theta,z))f(z)\textit{d}z\leq 1 \quad \textrm{and} \quad \int_{0}^{\infty}\mu_{2}(\theta,z))f(z)\textit{d}z\leq 1
\end{equation}
where $f(z)$ is the probability density function (pdf) of the power of the channel fading coefficients.

\subsection{Effective Capacity}

Wu and Negi in \cite{Wu} defined the effective capacity as the maximum constant arrival rate that
can be supported by a given channel service process while also satisfying a statistical QoS
requirement specified by the QoS exponent $\theta$. If we define $Q$ as the stationary queue
length, then $\theta$ is defined as the decay rate of the tail of the distribution of the queue
length $Q$:
\begin{equation}\label{decayrate}
\lim_{q\rightarrow \infty}\frac{\log P(Q\geq q)}{q}=-\theta.
\end{equation}
Hence, we have the following approximation for the buffer violation
probability for large $q_{max}$: $P(Q\geq q_{max})\approx e^{-\theta
q_{max}}$. Therefore, larger $\theta$ corresponds to more strict QoS
constraints, while the smaller $\theta$ implies looser constraints.
In certain settings, constraints on the queue length can be linked
to limitations on the delay and hence delay-QoS constraints. It is
shown in \cite{Liu} that $P\{D\geq d_{max}\}\leq c\sqrt{P\{Q\geq
q_{max}\}}$ for constant arrival rates, where $D$ denotes the
steady-state delay experienced in the buffer. In the above
formulation, $c$ is a positive constant, $q_{max}=ad_{max}$ and $a$
is the source arrival rate.

The effective capacity for a given QoS exponent $\theta$ is given by
\begin{equation}\label{exponent}
-\lim_{t\rightarrow \infty}\frac{1}{\theta t}\log_{e}\mathbb{E}\{e^{-\theta S(t)}\}=-\frac{\Lambda(-\theta)}{\theta}
\end{equation}
where $S(t)=\sum_{k=1}^{t}r(k)$ is the time-accumulated service
process, and $\{r(k),k=1,2,\dots\}$ is defined as the discrete-time,
stationary and ergodic stochastic service process. Note that the
service rate is $r(k) = r_{1}(k)(T-N)$ if the cognitive system is in
state 1 or 3 at time $k$. Similarly, the service rate is $r(k) =
r_{2}(k)(T-N)$ in state 4. In the OFF state, the determined
transmission rate exceeds the instantaneous channel capacity and
reliable communication is not possible. Therefore, the service rate
in this state is effectively zero.

In the next result, we provide the effective capacity for the
cognitive radio channel and state transition model described in the
previous section.

\begin{theo}\label{theo:variableratepower}
For the cognitive radio channel with the state transition model
given in Section \ref{subsec:state}, the normalized effective
capacity in bits/s/Hz is given by
\begin{align}\label{maximized}
&R_E(\tSNR,\theta)=\max_{\substack{\mu_{1}(\theta,z):\E_z\{\mu_{1}(\theta,z)\}\le 1 \\ \mu_{2}(\theta,z):\E_z\{\mu_{2}(\theta,z)\}\le 1}}\nonumber\\
-\frac{1}{\theta TB}\log_e\bigg[&\left(\rho P_{d}
+(1-\rho)P_{f}\right)\E_z\{e^{-(T-N)\theta r_{1}}\} \nonumber
\\
&+(1-\rho)(1-P_{f})\E_z\{e^{-(T-N)\theta
r_{2}}\}+\rho(1-P_{d})\bigg].
\end{align}
\end{theo}

\emph{Proof:} In \cite[Chap. 7, Example 7.2.7]{Performance}, it is
shown for Markov modulated processes that
\begin{gather} \label{eq:theta-envelope}
\frac{\Lambda(\theta)}{\theta} = \frac{1}{\theta} \log_e sp(\phi(\theta)R)
\end{gather}
where $sp(\phi(\theta)R)$ is the spectral radius (i.e., the maximum
of the absolute values of the eigenvalues) of the matrix
$\phi(\theta)R$, $R$ is the transition matrix of the underlying
Markov process, and $\phi(\theta) = \text{diag}(\phi_1(\theta),
\ldots, \phi_M(\theta))$ is a diagonal matrix whose components are
the moment generating functions of the processes in given states.
The rates supported by the cognitive radio channel with the state
transition model described in the previous section can be seen as a
Markov modulated process and hence the setup considered in
\cite{Performance} can be immediately applied to our setting. Since
the transmission rates are non-random and fixed in each state in the
cognitive channel, we can easily find that
\begin{align}\label{diago}
\phi(\theta)=&\text{diag}\bigg \{\E_z\{e^{-(T-N)\theta r_{1}}\},1\nonumber\\
&,\E_z\{e^{-(T-N)\theta r_{1}}\},\E_z\{e^{-(T-N)\theta r_{2}}\} \bigg\}.
\end{align}
Then, we have
\begin{center}
$\phi(\theta)R=\left[
\begin{array}{cccc}
\phi_{1}(\theta)p_{1} & . & . & \phi_{1}(\theta)p_{1} \\
. &   &   & . \\
. &   &   & . \\
\phi_{4}(\theta)p_{4} & . & . & \phi_{4}(\theta)p_{4} \\
\end{array}
\right]$.
\end{center}
Since $\phi(\theta)R$ is a matrix with unit rank, we can readily find that
\begin{align}
sp(\phi(\theta)R)&=\phi_{1}(\theta)p_{1} +...+ \phi_{4}(\theta)p_{4}
\\
&=(p_{1}+p_{3})\E_z\{e^{-(T-N)\theta r_{1}}\}\nonumber\\
&+ p_{2} + p_{4}\E_z\{e^{-(T-N)\theta r_{2}}\} \label{eq:sp}.
\end{align}

Then, combining (\ref{eq:sp}) with (\ref{eq:theta-envelope}) and (\ref{exponent}),
and noting that choice of the power policies $\mu_{1}(\theta,z)$ and
$\mu_{2}(\theta,z)$ can be optimized leads to the effective capacity formula
given in (\ref{maximized}). \hfill $\square$

Having obtained the expression for the effective capacity, we now
derive the optimal power adaptation strategies that maximize the
effective capacity.
\begin{theo} \label{theo:optimalpower}
The optimal power adaptation policies that maximize the effective
capacity are given by
\begin{equation}\label{mu1}
\mu_{1}(\theta,z)=\left\{
          \begin{array}{ll}
            \frac{1}{\tSNR_{1}}\left(\frac{1}{\gamma_1^{\frac{1}{a+1}}}\frac{1}{z^{\frac{a}{a+1}}}-\frac{1}{z}\right), & \hbox{$z>\gamma_1$} \\
            0, & \hbox{otherwise}
          \end{array}
        \right.
\end{equation}
and
\begin{equation}\label{mu2}
\mu_{2}(\theta,z)=\left\{
          \begin{array}{ll}
            \frac{1}{\tSNR_{4}}\left(\frac{1}{\gamma_{2}^{\frac{1}{a+1}}}\frac{1}{z^{\frac{a}{a+1}}}-\frac{1}{z}\right), & \hbox{$z>\gamma_{2}$} \\
            0, & \hbox{otherwise.}
          \end{array}
        \right.
\end{equation}
where $a=(T-N)B\theta/\log_e2$. $\gamma_{1}$ and $\gamma_{2}$ are
the threshold values in the power adaptation policies and they can
be found from the average power constraints in (\ref{mu}) through
numerical techniques.
\end{theo}

\emph{Proof:} Since logarithm is a monotonic function,  the optimal
power adaptation policies can also be obtained from the following
minimization problem
\begin{align}\label{objective}
\min_{\substack{\mu_{1}(\theta,z):\E_z\{\mu_{1}(\theta,z)\}\le 1 \\ \mu_{2}(\theta,z):\E_z\{\mu_{2}(\theta,z)\}\le 1}} \bigg\{&\left(\rho P_{d}
+(1-\rho)P_{f}\right)\E_z\{e^{-(T-N)\theta r_{1}}\}\nonumber\\&+(1-\rho)(1-P_{f})\E_z\{e^{-(T-N)\theta r_{2}}\}\bigg\}.
\end{align}
It is clear that the objective function in (\ref{objective}) is
strictly convex and the constraint functions in (\ref{mu}) are
linear with respect to $\mu_{1}(\theta,z)$ and $\mu_{2}(\theta,z)$.
Then, forming the Lagrangian function and setting the derivatives of
the Lagrangian with respect to $\mu_{1}(\theta,z)$ and
$\mu_{2}(\theta,z)$ equal to zero, we obtain
\begin{align}\label{lagran1}
&\left\{\lambda_{1}-\frac{a\tSNR_{1}z\left[\rho P_{d}+(1-\rho)P_{f}\right]}{\left[1+\mu_{1}(\theta,z)z\tSNR_{1}\right]^{a+1}}\right\}f(z)=0\\
&\left\{\lambda_{2}-\frac{a\tSNR_{4}z\left(1-\rho\right)\left(1-P_{f}\right)}{\left[1+\mu_{2}(\theta,z)z\tSNR_{4}\right]^{a+1}}\right\}f(z)=0 \label{lagran2}
\end{align}
where $\lambda_1$ and $\lambda_2$ are the Lagrange multipliers.
Defining $\gamma_{1}=\frac{\lambda_{1}}{\left[\rho
P_{d}+(1-\rho)P_{f}\right]a\tSNR_{1}}$ and
$\gamma_{2}=\frac{\lambda_{2}}{\left(1-\rho\right)\left(1-P_{f}\right)a\tSNR_{4}}$,
and solving (\ref{lagran1}) and \eqref{lagran2}, we obtain optimal
power policies
given in (\ref{mu1}) and (\ref{mu2}). 
\hfill $\square$

The optimal power allocation schemes identified in Theorem
\ref{theo:optimalpower} are similar to that given in \cite{zhang}.
However, in the cognitive radio channel, we have two allocation
schemes depending on the presence or absence of active primary
users. Note that the optimal power allocation in the presence of
active users, $\mu_{1}(\theta,z(i)) =
\frac{P_1(\theta,z(i))}{\overline{P}_1}$, has to be performed under
a more strict average power constraint since $\overline{P}_1 <
\overline{P}_2$. Note also that under certain fading conditions, we
might have $\mu_{1}(\theta,z(i)) > \overline{P}_1$, causing more
interference to the primary users. Therefore, it is also of interest
to apply only rate adaptation and use fixed-power transmission in
which case we have $\mu_{1}(\theta,z(i)) = \mu_{2}(\theta,z(i)) =
1$. We can immediately see from the result of Theorem
\ref{theo:variableratepower} that the effective capacity of
fixed-power/variable-rate transmission is
\begin{align}
R_E(\tSNR,\theta)= &-\frac{1}{\theta TB}\log_e \nonumber\\
\bigg[&\left(\rho P_{d}
+(1-\rho)P_{f}\right)\E_z\{e^{-(T-N)\theta r_{1}}\}\nonumber\\+&(1-\rho)(1-P_{f})\E_z\{e^{-(T-N)\theta r_{2}}\}+\rho(1-P_{d})\bigg]
\end{align}
where $r_1 = B\log_2(1+z\tSNR_{1})$ and $r_2 = B\log_2(1+z\tSNR_{4})$.

\section{Numerical Results}\label{sec:numericalresults}
\begin{figure}
\begin{center}
\includegraphics[width = \figsize\textwidth]{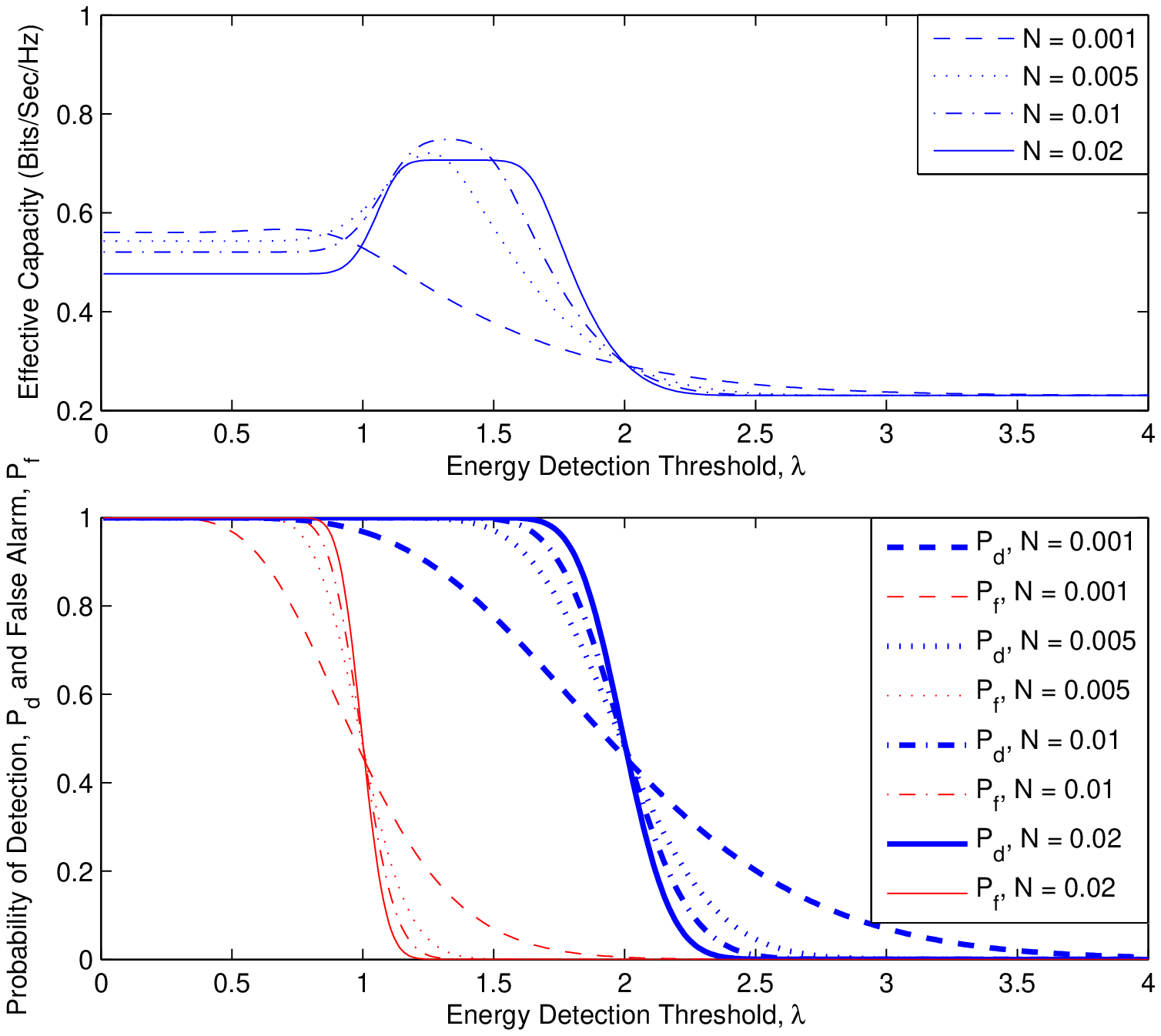}
\caption{Effective Capacity and $P_{f}$-$P_{d}$ vs. Channel Detection Threshold, $\lambda$. } \label{fig:fig2}
\end{center}
\end{figure}

In this section, we present the numerical results. In Figure
\ref{fig:fig2}, we plot the effective capacity with respect to energy
detection threshold $\lambda$ for different values of detection
duration $N$. We compare the probabilities of false alarm, $P_{f}$
and detection, $P_{d}$. The channel bandwidth is assumed to be
$10$kHz and the average input $SNR$ values, when the channel is
correctly detected, are $SNR_{1}=0$dB and $SNR_{4}=10$dB. The QoS
exponent $\theta=0.01$. The channel is assumed to be busy with an
average probability of $\rho=0.1$ and the duration of the block is
$T=0.1$s. As we see in Fig. \ref{fig:fig2}, $P_{d}$ and
$P_{f}$ depend on the duration of channel detection. With
increasing channel detection duration, we get more concrete $P_{d}$
and $P_{f}$ values. We can easily see that with a reasonable channel
detection duration, there is an optimal energy detection threshold
interval, which can be observed to be in between the values, 1.2 and     
1.7. In addition, we should note that there is also an optimal
channel detection duration. As we can see in Fig. \ref{fig:fig2},
when $N=0.01$s, we have higher effective capacity than we have when
$N=0.02$s. It is because the duration allocated for data transmission
is decreasing. The maximum effective capacity is $0.11$
bits/s/Hz, which is obtained when $N=0.01$s and $\lambda$ is
in between 1.2 and 1.7. With increasing energy
detection threshold, the effective capacity decreases sharply, which is
because the time allocated for the OFF state in which there is no reliable data transmission increases.

\begin{figure}
\begin{center}
\includegraphics[width = \figsize\textwidth]{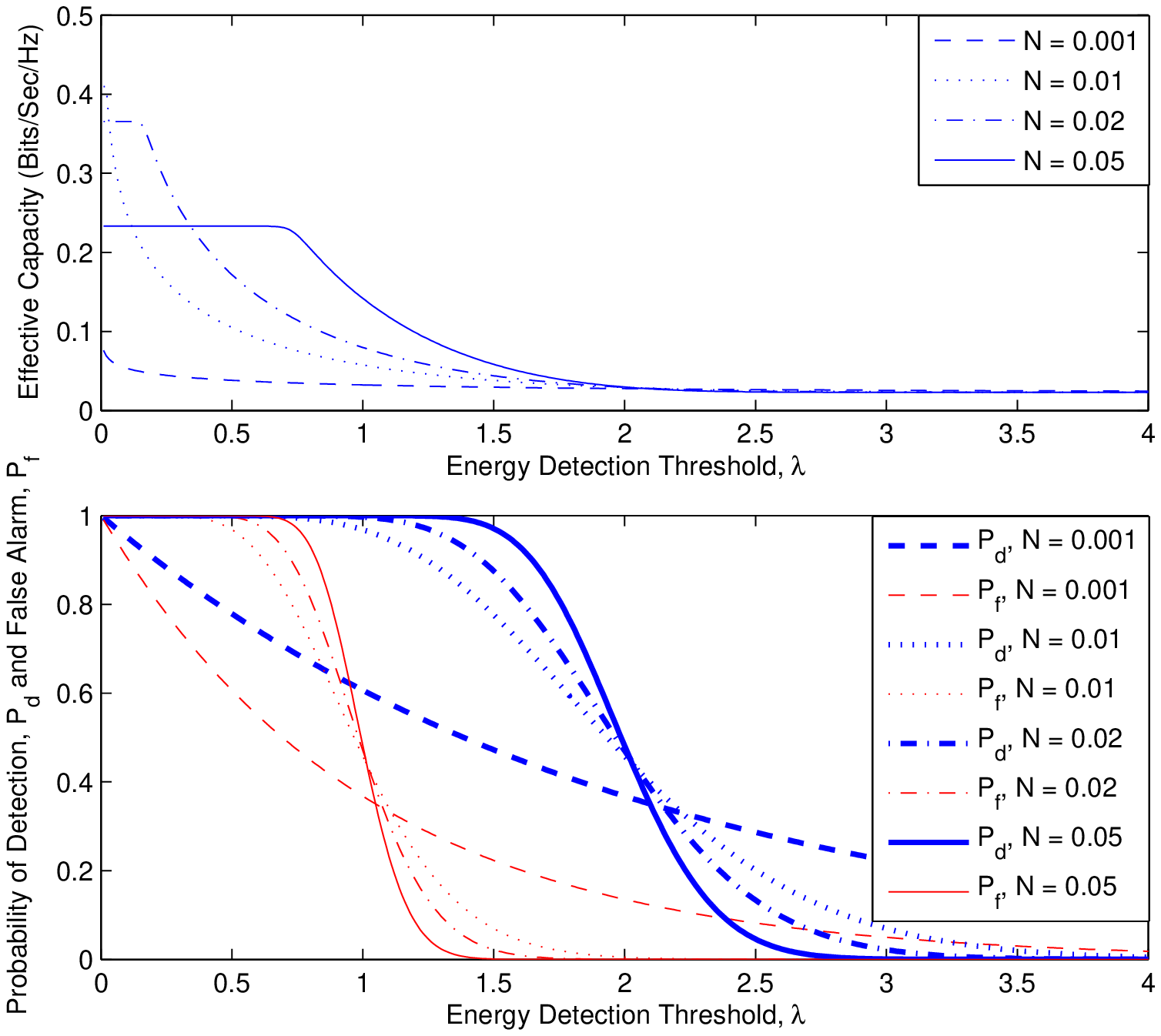}
\caption{Effective Capacity and $P_{f}$-$P_{d}$ vs. Channel Detection Threshold, $\lambda$. } \label{fig:fig3}
\end{center}
\end{figure}

In Fig. \ref{fig:fig3}, the input SNR values are kept the same as the
ones used in Fig. \ref{fig:fig2} while we change the QoS
exponent to $\theta=1$ and the bandwidth to $B=1$kHz. Again, the
block duration is $T=0.1$s. We consider 4 different detection
durations and compare the effective capacity results. As we can
easily realize, $P_{d}$ and $P_{f}$ show similar patterns as in Fig. \ref{fig:fig2}. But, since the bandwidth is small
compared to the one in Fig. \ref{fig:fig2}, the quality of channel
detection has decreased. Therefore, the highest effective capacity
rate is obtained not when $P_{d}=1$ and $P_{f}=0$. Hence, there is a
tradeoff between the effective capacity rate and disturbance caused to the
primary users.

\begin{figure}
\begin{center}
\includegraphics[width = \figsize\textwidth]{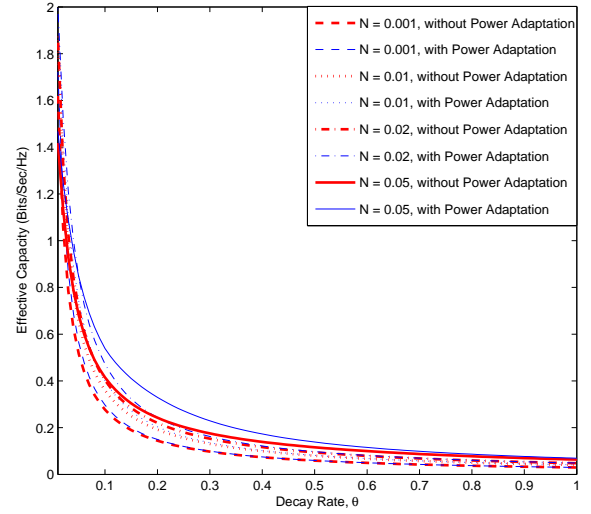}
\caption{Effective Capacity vs. Decay Rate, $\theta$. }
\label{fig:fig4}
\end{center}
\end{figure}

In Fig. \ref{fig:fig4}, we plot the effective capacity with respect to
decay rate, $\theta$, where we consider the same input SNR values as
we have in Fig. \ref{fig:fig2} and Fig. \ref{fig:fig3}.
The channel bandwidth is $B=1$kHz, block duration is $T=0.1$s, and the energy detection threshold is $\lambda=1.4$. We
obtained the effective capacity both with and without optimizing the
transmission power adaptation policy. For a given channel detection duration, the effective
capacity is decreasing as $\theta$ increases, as expected. On the other hand, for fixed
$\theta$ and $\lambda$ values, we obtain different optimum
channel detection duration values. For instance, for lower decay
rates $\theta$, we obtain higher effective capacity when $N=0.02$s. For
higher decay rate values, choosing $N=0.05$s provides higher effective capacity values. We should also note
that since $P_{d}$ and $P_{f}$ are independent of the decay rate,
for each detection duration there exist particular $P_{d}$ and
$P_{f}$ values and they are constant for all decay rates. Finally,
we can notice that for high values of $\theta$, power adaptation
does not provide much gain in terms of the effective capacity.

\section{Conclusion}\label{sec:conclusion}
In this paper, we have analyzed the effective capacity of cognitive
radio channels in order to identify the performance levels and to
determine the interactions between throughput and channel sensing
parameters in the presence of QoS constraints. We have assumed that both the receiver and the
transmitter have CSI. We have initially
constructed a 4-state-transition model for cognitive transmission
and then obtained expressions for the effective capacity.  This
analysis is conducted for variable transmission rates and variable
transmission powers.  Through numerical results, we have
investigated the impact of channel sensing duration and threshold,
detection and false alarm probabilities, and QoS limitations on the
throughput. Several insightful observations are made. For instance, we have seen that
the channel sensing duration and threshold have great impact on the effective
capacity.  Also, we have noted that the gains provided by power adaptation diminishes as $\theta$ increases.


\vspace{-0.2cm}
\begin{thebibliography}{99}

\bibitem{Performance} C.-S. Chang, Performance Guarantees in Communication Networks,
\emph{New York: Springer, 1995.}

\bibitem{Poor-book} H. V. Poor, An Introduction to Signal Detection and Estimation, 2nd ed.,
\emph{Springer-Verlag, 1994.}

\bibitem{liang} Y.-C. Liang, Y. Zheng, E. C. Y. Peh, and A. T. Hoang, ``Sensing-throughput tradeoff for cognitive radio networks,"
\emph{IEEE Trans. Wireless Commun., vol. 7, no. 4, pp. 1326-1337, Apr. 2008.}

\bibitem{Wu} D. Wu and R. Negi, ``Effective Capacity: A Wireless Link Model for Support of Quality of Service,"
\emph{IEEE Trans. Wireless Commun., vol. 2, no. 4, pp. 630-643. July 2003.}

\bibitem{Liu} L. Liu, and J.-F. Chamberland, ``On The Effective Capacities of Multiple-Antenna Gaussian Channels,"
\emph{IEEE International Symposium on Information Theory, Toronto, 2008.}

\bibitem{zhang} J. Tang and X. Zhang, ``Quality-of-Service Driven Power and Rate Adaptation Over Wireless Links,"
\emph{IEEE Trans. Wireless Commun., vol. 6, no. 8, pp. 3058-3068, Aug. 2007.}

\bibitem{Geirhofe} S. Geirhofer, L. Tong, and B.M. Sadler, ``A Measurement-Based Model for Dynamic Spectrum Access in WLAN Channels,"
\emph{Military Communications Conference, Washington D.C., 2006.}

\bibitem{Lai} H. Jiang, L. Lai, R. Fan, and H.V. Poor, ``Cognitive Radio: How to Maximally Utilize Spectrum Opportunities in Sequential Sensing,"
\emph{Global Telecommunications Conference, New Orleans, LA, USA, Nov. 30-Dec. 4, 2008.}

\bibitem{Srinivasa} S. Srinivasa, S.A. Jafar, and N. Jindal, ``On the Capacity of the Cognitive Tracking Channel,"
\emph{IEEE International Symposium on Information Theory, Seattle, Jul. 9-14, 2006.}

\bibitem{Gursoy} M.C. Gursoy, D. Qiao, and S. Velipasalar, ``Analysis of Energy Efficiency in Fading Channel under QoS Constraints,"
\emph{IEEE Global Communications Conference, New Orleans, LA, USA, Nov. 30-Dec. 4, 2008.}

\bibitem{Liu2} L. Liu, P. Parag, J. Tang, W.-Y. Chen and J.-F. Chamberland, ``Resource Allocation and Quality of Service Evaluation for Wireless Communication Systems Using Fluid Models,"
\emph{IEEE Trans. Inform. Theory, vol. 53, no. 5, pp. 1767-1777, May 2007.}

\bibitem{Tang} J. Tang and X. Zhang, ``Quality-of-Service Driven Power and Rate Adaptation Over Wireless Links,"
\emph{IEEE Trans. Wireless Commun., vol. 6, no. 8, pp. 3058-3068, Aug. 2007.}

\bibitem{Sami} S. Akin and M.C. Gursoy, ``Effective Capacity Analysis of Cognitive Radio Channels for Quality of Service Provisioning,"
\emph{Global Telecommunications Conference, Honolulu, Hawaii, USA, Nov. 30-Dec. 4, 2009.}


\end{thebibliography}
\end{document}